\input harvmac
\input amssym
\input epsf

\newcount\figno
\figno=0 
\def\fig#1#2#3{
\par\begingroup\parindent=0pt\leftskip=1cm\rightskip=1cm\parindent=0pt
\baselineskip=11pt
\global\advance\figno by 1
\midinsert
\epsfxsize=#3
\centerline{\epsfbox{#2}}
\vskip 12pt
{\bf Fig.\ \the\figno: } #1\par
\endinsert\endgroup\par
}
\def\figlabel#1{\xdef#1{\the\figno}}

\def\wt{\widetilde{\lambda}}
\def\CP{\Bbb{P}}

\lref\ManganoXK{
M.~L.~Mangano, S.~J.~Parke and Z.~Xu,
``Duality And Multi-Gluon Scattering,''
Nucl.\ Phys.\ B {\bf 298}, 653 (1988).
}

\lref\BerendsME{
F.~A.~Berends and W.~T.~Giele,
``Recursive Calculations For Processes With N Gluons,''
Nucl.\ Phys.\ B {\bf 306}, 759 (1988).
}

\lref\KleissNE{
R.~Kleiss and H.~Kuijf,
``Multi - Gluon Cross-Sections And Five Jet Production At Hadron Colliders,''
Nucl.\ Phys.\ B {\bf 312}, 616 (1989).
}

\lref\BerendsZN{
F.~A.~Berends and W.~T.~Giele,
``Multiple Soft Gluon Radiation In Parton Processes,''
Nucl.\ Phys.\ B {\bf 313}, 595 (1989).
}

\lref\ManganoBY{
M.~L.~Mangano and S.~J.~Parke,
``Multiparton Amplitudes In Gauge Theories,''
Phys.\ Rept.\  {\bf 200}, 301 (1991).
}

\lref\DixonWI{
L.~J.~Dixon,
``Calculating scattering amplitudes efficiently,''
arXiv:hep-ph/9601359.
}

\lref\BernJE{
Z.~Bern, L.~J.~Dixon and D.~A.~Kosower,
``Progress in one-loop QCD computations,''
Ann.\ Rev.\ Nucl.\ Part.\ Sci.\  {\bf 46}, 109 (1996)
[arXiv:hep-ph/9602280].
}

\lref\Stanley{
R.~P.~Stanley,
``Hipparchus, Plutarch, Schroeder, and Hough,''
{\it The American Mathematical Monthly}, Vol.~104, No.~4,
344 (April 1997).
}

\nref\WittenNN{
E.~Witten,
``Perturbative gauge theory as a string theory in twistor space,''
arXiv:hep-th/0312171.
}

\nref\RoibanVT{
R.~Roiban, M.~Spradlin and A.~Volovich,
``A googly amplitude from the B-model in twistor space,''
arXiv:hep-th/0402016.
}

\nref\BerkovitsHG{
N.~Berkovits,
``An alternative string theory in twistor space for N = 4 super-Yang-Mills,''
arXiv:hep-th/0402045.
}

\nref\RoibanKA{
R.~Roiban and A.~Volovich,
``All googly amplitudes from the B-model in twistor space,''
arXiv:hep-th/0402121.
}

\nref\NeitzkePF{
A.~Neitzke and C.~Vafa,
``N = 2 strings and the twistorial Calabi-Yau,''
arXiv:hep-th/0402128.
}

\nref\CachazoKJ{
F.~Cachazo, P.~Svrcek and E.~Witten,
``MHV vertices and tree amplitudes in gauge theory,''
arXiv:hep-th/0403047.
}

\nref\ZhuKR{
C.~J.~Zhu,
``The googly amplitudes in gauge theory,''
arXiv:hep-th/0403115.
}

\nref\NekrasovJS{
N.~Nekrasov, H.~Ooguri and C.~Vafa,
``S-duality and Topological Strings,''
arXiv:hep-th/0403167.
}

\nref\BM{
N.~Berkovits and L.~Motl,
``Cubic Twistorial String Field Theory'',
arXiv:hep-th/0403187.
}

\Title{\vbox{\baselineskip12pt
	\hbox{hep-th/0403190}
	\hbox{NSF-KITP-04-35}
}}{On the Tree-Level $S$-Matrix of Yang-Mills Theory}

\centerline{
Radu Roiban${}^\dagger$,
Marcus Spradlin${}^\ddagger$ and
Anastasia Volovich${}^\ddagger$}

\bigskip
\bigskip

\centerline{${}^\dagger$Department of Physics, University of California}
\centerline{Santa Barbara, CA 93106 USA}

\smallskip

\centerline{${}^\ddagger$Kavli Institute for Theoretical Physics}
\centerline{Santa Barbara, CA 93106 USA}

\bigskip
\bigskip

\centerline{\bf Abstract}

\bigskip

In this note we further investigate the procedure for computing tree-level
amplitudes in Yang-Mills theory from connected instantons in the B-model on
$\CP^{3|4}$, emphasizing that the problem of calculating Feynman diagrams is
recast into the problem of finding solutions to a certain set of algebraic
equations.  We show that the B-model correctly reproduces all 6-particle
amplitudes, including non-MHV amplitudes with three negative and three
positive helicity gluons.  As a further check, we also show that $n$-particle
amplitudes obtained from the B-model obey a number of properties required of
gauge theory, such as parity symmetry (which relates an integral over degree
$d$ curves to one over degree $n-d-2$ curves) and
the soft and collinear gluon poles.

\Date{March 2004}

\listtoc
\writetoc

\newsec{Introduction}

In \WittenNN\ Witten proposed a remarkable
connection between scattering amplitudes in Yang-Mills (YM) theory
and a certain topological string theory, the B-model
on $\CP^{3|4}$ (recent related work includes
\refs{\RoibanVT - \BM}).
This conjecture leads to the following formula (equivalent
to one first written
down in \RoibanVT\ and studied further
in \refs{\BerkovitsHG,\RoibanKA}) for the $n$-particle amplitude,
written in a manifestly ${\cal{N}}=4$ supersymmetric notation:
\eqn\answer{
\eqalign{
A_n= i (2 \pi)^4 g_{\rm YM}^{n-2} \sum_{d=1}^{n-3}
\int
d{\cal{M}}_{n,d}
\prod_{i=1}^n \delta^2 \left(
\lambda_i^\alpha-\xi_i P_i^\alpha
\right)
\prod_{k=0}^d \delta^2 \left(
\sum_{i=1}^n \xi_i \sigma_i^k \wt_i^{\dot{\alpha}}
\right) \delta^4 \left(
\sum_{i=1}^n \xi_i \sigma_i^k \eta_{iA} \right).
}}
The details of this formula will be clarified in section 2,
but we have written it down here in order
to stress its simplicity and importance.
We believe that \answer\ encapsulates
the {\it complete} $n$-particle tree-level $S$-matrix of YM theory
(for any gauge group), thereby providing an exact solution
of classical YM theory in four dimensions.
This formula sums up a huge
number of Feynman diagrams (see for
example Fig.~1) into an expression which
fits on a single line.
In this note we provide strong evidence supporting our
confidence in this formula
and explore some of its structure.

\fig{The standard computation
of a six-gluon tree amplitude
requires summing 220 Feynman diagrams
(in conventional gauges) \KleissNE.}
{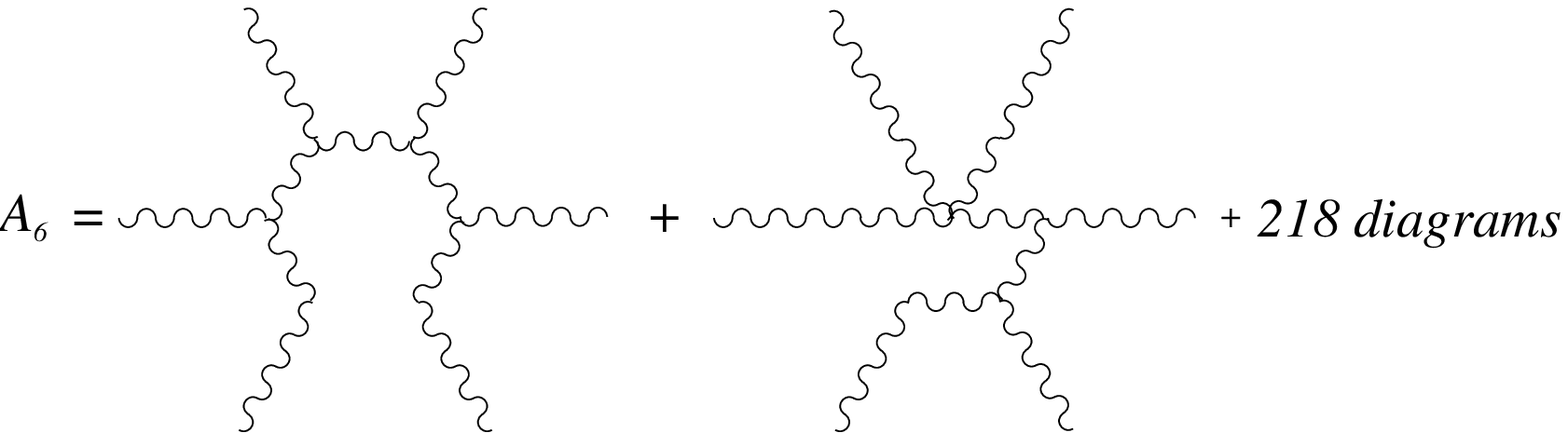}{4.0in}

The formula \answer\ was derived by considering
the contribution to the scattering amplitude from a single
connected instanton (= holomorphic
curve in $\CP^{3|4}$) of degree $d$ in the topological B-model.
(As explained in \WittenNN, counting the
fermionic modes reveals that the degree $d$ is related
to the total helicity $\sum h_i$ of all $n$ particles according
to $d = \ha (n - \sum h_i - 2)$.)
In \WittenNN\ Witten speculated that one might have to consider, in
addition to \answer,
contributions from collections of disconnected
instantons of degrees $d_i$ with $\sum d_i = d$.  (See Fig.~2
for a schematic depiction for the $n=6$, $d=3$ amplitude.)

However, it was found in \refs{\RoibanVT,
\RoibanKA} that the formula \answer\ correctly reproduces
the known YM result for
the mostly minus MHV (maximally-helicity violating)
amplitudes (sometimes called `googly') that are related to the
mostly plus MHV
amplitudes by complex conjugation.
Even though the googly amplitudes are
calculated from an integral
over the moduli space of instantons of arbitrarily high degree,
precise agreement was found with gauge theory without the
need for additional contributions.

More recently, a novel method for calculating YM tree amplitudes, also
motivated by the B-model on $\CP^{3|4}$, was
proposed in a very interesting paper \CachazoKJ.
The starting point for their proposal involved considering
only completely
disconnected instantons (i.e., $d$ instantons of degree 1).
Remarkably, it was found that their rule also gives
correct gauge theory amplitudes.
The B-model seems to give two separately correct methods for
calculating YM tree amplitudes, rather than a set of contributions
which need to be summed (see Fig.~2).

\fig{Schematic depiction of how one might have thought to
organize
the calculation of the
6-particle mostly minus MHV ($d=3$) amplitude in the B-model
on $\CP^{3|4}$.
The dark $\times$'s mark the insertions of the 6 external
particles, the dotted line is a twistor space propagator (constructed
in \CachazoKJ),
and the solid lines represent instantons (i.e., holomorphic
curves in $\CP^{3|4}$) of degree $d=1,2,3$ (schematically encoded in
the waviness of the curve).  Although one might have expected
that it would be necessary to sum together
contributions of all three types,
we find  that the {\it single} diagram
of the first type (studied here and in \refs{\RoibanVT,\RoibanKA}),
and the sum of the 21 diagrams of the third type
(studied in \CachazoKJ), {\it separately} give the correct
gauge theory answer.}
{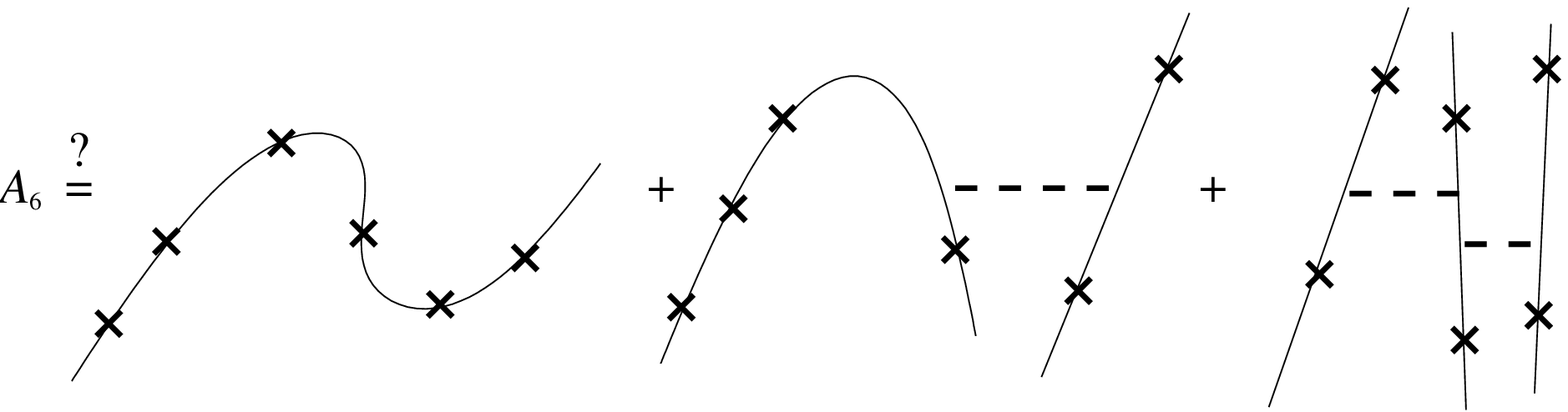}{4.5in}

The proposal of \CachazoKJ\ involves a diagrammatic expansion
which bears no apparently obvious
connection to the formula \answer, except that they both
seem to be correct. 
It would be very interesting to understand
directly the relation between these two methods.
Moreover,
if the B-model on $\CP^{3|4}$ gives us two not obviously
equivalent formulas for YM amplitudes, then it will likely give
us an infinite family of formulas (which roughly speaking
weight the different types of diagrams in Fig.~2 differently).
Undoubtedly we have only encountered the tip of the iceberg
connecting the topological B-model to Yang-Mills
amplitudes.

So far the formula \answer\ had only been checked
for MHV and googly amplitudes \refs{\RoibanVT,\RoibanKA}.
In section 3 of this paper we confirm
that the formula also gives the correct 6-particle non-MHV amplitudes.
In section 4 we check that for any $n$ and $d$, \answer\ satisfies
a number of properties required of general Yang-Mills amplitudes,
such as the soft and collinear gluon limits.
Of particular importance is parity symmetry, which requires
that \answer\ should be invariant under $\lambda \leftrightarrow
\widetilde{\lambda}$.  This non-manifest symmetry of \answer\ is
proven explicitly in section 4.2 below.
We conclude with a list of open questions and puzzles.
First, however, we turn our attention to the details of \answer\ and
highlight a crucial fact about the formula:  namely, that it is not
really an integral at all.

\newsec{The Main Formula}

In this section we first clarify the ingredients
appearing in the formula \answer\ 
and then investigate some of its mathematical properties.
The quantity $A_n$ in \answer\ denotes the color-stripped
$n$-particle partial amplitude (see for example \DixonWI),
and we employ the spinor helicity notation in writing
$A_n$ as a function
of $(\lambda^\alpha_i, \wt^{\dot{\alpha}}_i, \eta_{iA})$, $i=1,\ldots,n$,
where $\lambda$ and $\wt$ are 
commuting 
real two-component spinors of positive and negative
chirality,
respectively\foot{For simplicity we work in signature
$+$ $+$ $-$ $-$, where such spinors are possible.  It is straightforward
to analytically continue the tree-level YM amplitudes to signature
$+$ $-$ $-$ $-$ if desired.}, and $\eta_A$ is the four-component
Grassmann coordinate of ${\cal{N}} = 4$ superspace.

The $P^\alpha$ are two degree $d$ polynomials in $\sigma$
which we parametrize as
\eqn\pdef{
P^\alpha_i  = \sum_{k=0}^d a^\alpha_k \sigma_i^k
}
in terms of $2d + 2$ coefficients (moduli) $a^\alpha_k$.
When needed, we
will follow the conventions of \refs{\RoibanVT,\RoibanKA}
in denoting $P^1_i = A_i$ and $P^2_i = B_i$.
The measure for integration in \answer\ is
\eqn\measure{
d{\cal{M}}_{n,d} = {d^{2d+2} a\, d^n \sigma\,d^n \xi
\over {\rm vol}({\rm GL(2)})}
\prod_{i=1}^n {1 \over \xi_i (\sigma_i - \sigma_{i+1})}.
}
The factor of $1/{\rm vol}({\rm GL(2)})$ is included because the
integrand is invariant under a certain GL(2) symmetry and
so the integral
would otherwise be infinite.  Practically, the consequence
of this factor is simply that we can choose to fix four of the variables
(any one of the $a$'s and any three of the $\sigma$'s)
at the expense of introducing a Jacobian factor of
\eqn\aaa{
J = a (\sigma_i - \sigma_j) (\sigma_j - \sigma_k)(\sigma_k - \sigma_i).
}
The choice of which $a$ and which three $\sigma$'s to
leave un-integrated is arbitrary and does not affect the final result.

\subsec{A key point}

The single most important fact about the integral \answer\ is
that it is not really an integral.
To see this, let us start by
showing that \answer\ respects momentum conservation.
Taking a particular linear combination of the quantities
set to zero by the delta functions  gives
\eqn\pcon{
0 = \sum_{k=0}^d a_k^\alpha \left( \sum_{i=1}^n
\xi_i \sigma_i^k \wt_i^{\dot{\alpha}} \right)
= \sum_{i=1}^n \xi_i P_i^\alpha \wt_i^{\dot{\alpha}}
= \sum_{i=1}^n \lambda_i^\alpha \wt_i^{\dot{\alpha}}
= \sum_{i=1}^n p_i^{\alpha \dot{\alpha}},
}
where we used the definition \pdef\ and some more delta
functions from \answer.  Therefore, the delta functions in \answer\ 
indeed force overall momentum conservation.

At the practical level, this means we can `pull out' the overall
factor of $\delta^4(\sum p_i)$
at the expense of introducing a Jacobian, by using an identity such as
\eqn\aaa{\eqalign{
&\prod_{i=1}^n \delta \left( {\lambda_i^2 \over \lambda_i^1}
- {B_i \over A_i} \right) \prod_{k=0}^d
 \delta^2 \left( \sum_{i=1}^n
{ \wt_i^{\dot{\alpha}} \lambda_i^1 \sigma_i^k \over A_i}
\right)
\cr
&\qquad\qquad\qquad
= A_1 A_2\, \delta^4\left(\sum_{i=1}^n p_i\right)
\prod_{i=3}^n \delta \left( { \lambda_i^2 \over
\lambda_i^1} - {B_i \over A_i}\right) \prod_{k=1}^d
\delta^2 \left( \sum_{i=1}^n {\wt_i^{\dot{\alpha}} \lambda_i^1
\sigma_i^k \over A_i} \right)
}}
(where we used $\xi_i = \lambda_i^1/A_i$). 
In writing this identity we have made a particular choice of which
four delta functions to pull out.
There is however no canonical choice, and 
different choices are useful for different calculations
(and lead to different Jacobians), so it is convenient to leave
momentum conservation slightly scrambled into the delta functions
in \answer.
Note that supermomentum conservation $\delta^8 \left(\sum
\lambda_i^{\alpha} \eta_{iA}\right)$ pulls out similarly.

Let us now return to the claim that \answer\ is not really an integral.
The measure $d {\cal{M}}_{n,d}$ in \measure\ has
$(2 d + 2) + (n ) + (n) - (4) = 2 n + 2 d - 2$ integration
variables, while the integrand in \answer\ has
$2 n + 2 d + 2$ delta functions.
If we `pull out' the overall momentum conservation delta functions, then
for any $n$ and $d$ there are precisely as many integration variables
as delta functions.
Therefore the entire integral is supported on a discrete set of points,
and the formula \answer\ is just
a recipe
to solve the $2 n + 2 d + 2$ polynomial equations
\eqn\equations{\eqalign{
\lambda_i^\alpha &= \xi_i \sum_{k=0}^d \sigma_i^k a^\alpha_k, \qquad
\alpha = 1,2, ~~ i=1,\ldots,n,\cr
0 &= \sum_{i=1}^n \xi_i \sigma_i^k \wt_i^{\dot{\alpha}}, \qquad
\dot{\alpha} = 1,2, ~~ k=0,\ldots d
}}
for the variables $(a^\alpha_k, \sigma_i, \xi_i)$, and then to sum a
certain Jacobian (obtained in the usual way from \answer) over
the collection of roots.

One of the most interesting questions about the system \equations\ is:
what is the number of roots $N_{n,d}$ for general $n$ and $d$?
At this point all we know for sure is that
\eqn\nroots{
N_{n,1} = N_{n,n-3} = 1, \qquad N_{6,2} = 4.
}
The first two cases are MHV and googly amplitudes
previously studied
in the literature, and $N_{6,2}$
is the non-MHV 6-particle amplitude
discussed in the following section.
In section 4 we prove that $N_{n,n-d-2}=N_{n,d}$.
Certainly it would be very interesting to have a better understanding
of the mathematics underlying the equations \equations.
In particular, it would be especially interesting to learn how
$N_{n,d}$ grows with $n$ and $d$.

\subsec{A complex puzzle}

A priori, the moduli $a_{k}^\alpha$ of the curve and the 
coordinates $\sigma_i$ on $\CP^1$ should all be complex variables.
In order to evaluate the integral \answer\ it is necessary
to specify an integration contour in this $2n + 2 d  - 2$
complex dimensional space.  In spacetime signature $+$ $+$ $-$ $-$
it makes sense to take $\lambda$ and $\wt$ to be independent
real variables, and it is natural to choose the 
integration contour in which all of the $a_k^\alpha$
and $\sigma_i$ are real.

For both the MHV ($d=1$) and googly ($d=n-3$) cases, the unique
root of the equations \equations\ indeed has the property
that $\sigma_i$ and $a_k^\alpha$ are real.
However, for the 6-particle amplitude with $d=2$, which we
discuss in section 3, there is a puzzle.  Depending on the
choice of $\lambda$ and $\wt$, there can be four real
roots, two real roots and one complex conjugate pair, or
two complex conjugate pairs.  The YM tree amplitude,
which is always real (forgetting the $i$ in front of \answer),
is reproduced only if all four roots are summed over, regardless
of whether they are real or complex.

The lesson from this analysis is that
restricting \answer\ to the contour where all $a$'s
and $\sigma$'s are real does not give
the correct gauge theory scattering amplitudes.
In fact, we do not know how to write any contour which makes
the integral formula \answer\ valid for arbitrary choices
of $\lambda$ and $\wt$.
This amplifies the comment we made at the beginning
of the previous subsection:  the formula \answer\ is not
really an integral.
To overcome this problem we
avoid thinking about \answer\ as an honest integral, but instead
view it as a recipe for finding the solution (which in
general can be complex) of \equations\ and then summing a
Jacobian over the set of roots.

\subsec{A diagrammatic expansion?}

 From this new standpoint,
let us ask ourselves whether the formula might have another, more
natural interpretation.
The fact that the computation of a scattering amplitude from
the formula \answer\ reduces to summing a certain quantity over
a finite set of points is reminiscent of some sort of diagrammatic
expansion, where, for example, \nroots\ suggests that
there is a single diagram for mostly plus and mostly minus MHV
amplitudes, 
while four diagrams contribute
to the 6-point non-MHV amplitudes.

It is tempting to wonder whether there is any connection between
such `diagrams' and the new diagrammatic expansion for YM
scattering amplitudes which was recently
proposed in \CachazoKJ.  According to their proposal,
$A_{n,d}$ is associated with the collection of
trees with $n$ cyclically labeled external legs and $d$ vertices, such
that 
each vertex has at least 3 legs.  For general $n$ and $d$
there are ${1 \over d} {n - 3 \choose d - 1}{n+d-2 \choose d-1}$
such graphs\foot{The counting of these graphs is equivalent to
a combinatorial problem which appeared in Plutarch's biographical
notes on Hipparchus \Stanley.  We are grateful to C.~Herzog
for many fun and enlightening discussions regarding
the combinatorics.}, which in all cases
except the trivial case $d=1$ is larger than \nroots.
(We have written the number of diagrams in ${\cal{N}}=4$
superspace.
For particular choices of helicities of the external particles
there are frequently fewer diagrams.)

However, the diagrams of \CachazoKJ\ have an additional
symmetry in the form of an arbitrary spinor
$\eta^{\dot{\alpha}}$ which drops out only after summing
together all of the graphs.
The number of diagrams is not gauge invariant, and special
choices of $\eta$ can set whole classes of diagrams to zero.
In contrast, our `diagrams' have no residual manifest
symmetry --- the GL(2) cancels out diagram by diagram (root by root)
and does
not change their number.  Maybe there is some choice of $\eta$
for which the diagrams of \CachazoKJ\ reduce, in number and
in value, to the contributions obtained from the roots of our
formula \answer?

We believe it is more likely
that the topological B-model has some huge symmetry group 
which relates
the formula \answer, with its associated `diagrams',
to the diagrammatic expansion of \CachazoKJ.  Their
parameter $\eta$ is a small residue of that huge symmetry.

\newsec{The 6-Particle Non-MHV Amplitudes}

In the previous section we introduced the formula \answer\ and
discussed its basic properties.  But what is the connection
between \answer\ and the $n$-particle scattering amplitude
in gauge theory?   In \WittenNN\ it was shown that
a prescription equivalent to the $d=1$ case of \answer\ reproduces
the mostly plus MHV
amplitudes in YM theory.
In \refs{\RoibanVT,\RoibanKA} it was shown that the formula
also works for mostly minus MHV amplitudes (sometimes called
googly or ${\overline{\rm MHV}}$).
These have
$d = n - 3$ and are related (in Minkowski
signature) to MHV amplitudes by complex conjugation.

Although the latter check involved an apparently nontrivial
integral over the moduli space of curves of arbitrary degree
in $\CP^{3|4}$, the question of whether \answer\ is correct
for genuinely non-MHV amplitudes was left open.
The simplest amplitudes which are neither MHV nor
googly are those with $n=6$ particles and $d=2$.
Since we work in a manifestly ${\cal{N}} = 4$ formalism, our
results apply simultaneously
to all possible helicity orderings
(when all six particles are gluons, there are three
independent helicity orderings:
 $+$~$+$~$+$~$-$~$-$~$-$,
$+$~$+$~$-$~$+$~$-$~$-$ and
$+$~$-$~$+$~$-$~$+$~$-$).

In this paper we report that the formula \answer, in the case
$n=6$ and $d=2$, precisely matches the 6-gluon scattering
amplitudes first
computed by Mangano, Parke and Xu \ManganoXK.
We originally obtained this result numerically, by
(1) choosing at random a collection of ($\lambda_i,\wt_i$)
(subject to overall momentum conservation \pcon),
(2) numerically solving the polynomial equations \equations,
which were always observed to have four roots, and then (3)
summing the Jacobian obtained from \answer\ over the four roots.
The whole calculation takes only a few seconds on a fast computer
and can be repeated as often as desired for different
$(\lambda_i,\wt_i)$. 
The result was always found to agree spectacularly
with the formula given
in \ManganoXK.  Note that all three independent helicity
configurations can be checked at the same time since the choice
of helicities only affects the fermion determinant and does not
change the value of the roots.

The only puzzle we encountered
is that occasionally, for some $(\lambda_i,\wt_i)$,
the roots are complex, as we discussed in subsection 2.2.
Precise agreement with gauge theory was nevertheless always found
by doing the most naive thing possible and summing over all four
roots, whether real or complex.

Unfortunately, it seems rather difficult to construct an analytic
proof that the formula \answer\ is correct for the case $n=6$,
$d=2$.  Let us now outline the best line of attack that we know
of at the moment.  We will not give precise formulas for
each intermediate step because they are extremely lengthy and
moreover because we are hopeful that a more clever
way of analyzing the equations will become available.
We believe that only after
the mathematical structure of the equations \equations\ is
better understood  (for arbitrary $n$ and $d$) will it be clear
how best to organize this calculation analytically.

\subsec{Constructing a Groebner basis: a sketch}

The most interesting result of the numerical analysis is
that the number of roots is $N_{6,2} = 4$, which does not
appear obvious from \equations.  
Recall that we can fix one of the $a$'s and three
of the $\sigma$'s (say $\sigma_1$, $\sigma_2$ and $\sigma_3$)
using the GL(2) symmetry.  The remaining $2n+2d-2 = 14$ `integration
variables' are fixed by solving \equations.
In fact, it turns out to be possible to express all of
the $a$'s, $\xi$'s and two of the remaining three $\sigma$'s as
rational functions of the final $\sigma$ (say $\sigma_6$).
Moreover, one can extract from \equations\ a single equation
which is quartic in $\sigma_6$
and does not depend on any of the other `integration
variables.'
The coefficients of this quartic polynomial are themselves
polynomials in $\sigma_1$, $\sigma_2$ and $\sigma_3$ and the
covariant kinematic quantities $[i\,j]$ and $\langle i\,j\rangle$.
The four roots of this master quartic equation determine
the solutions for all $14$ variables.

Here is a schematic description of how to derive this quartic
equation.
Choose some subset $S$ of the equations and use them to solve
for the $a$'s and $\xi$'s in terms of the $\sigma$'s.  Plugging the solution
into the remaining equations gives polynomial equations just
on the $\sigma$'s.  This process can be repeated many times by
starting with different sets $S$ of equations, leading to
a large number of polynomial
equations  on $\sigma_4$, $\sigma_5$ and $\sigma_6$.  The
game then is to find the common roots of these polynomial
equations.
In mathematical language, we need
to construct a Groebner basis for the ideal generated
by these polynomials.  Let us now be a little more specific.

Start with the equations on the top line of \equations.
By eliminating $\xi_i$ between the $\alpha=1$ and $\alpha=2$
versions of this equation, one arrives at
the six equations
\eqn\five{
\lambda_i^2 \sum_{k=0}^2 a_k^1 \sigma_i^k = \lambda_i^1
\sum_{k=0}^2 a_k^2 \sigma_i^k, \qquad
i = 1,\ldots,6,
}
which are conveniently expressed in matrix notation as
\eqn\bigone{
\pmatrix{
\lambda^1_1  & \lambda^1_1 \sigma_1 & \lambda^1_1 \sigma_1^2
& \lambda_1^2 & \lambda_1^2 \sigma_1 & \lambda_1^2 \sigma_1^2 \cr
\lambda^1_2  & \lambda^1_2 \sigma_2 & \lambda^1_2 \sigma_2^2
& \lambda_2^2 & \lambda_2^2 \sigma_2 & \lambda_2^2 \sigma_2^2 \cr
\lambda^1_3  & \lambda^1_3 \sigma_3 & \lambda^1_3 \sigma_3^2
& \lambda_2^2 & \lambda_2^2 \sigma_2 & \lambda_2^2 \sigma_3^2 \cr
\lambda^1_4  & \lambda^1_4 \sigma_4 & \lambda^1_4 \sigma_4^2
& \lambda_4^2 & \lambda_4^2 \sigma_4 & \lambda_4^2 \sigma_4^2 \cr
\lambda^1_5  & \lambda^1_5 \sigma_5 & \lambda^1_5 \sigma_5^2
& \lambda_5^2 & \lambda_5^2 \sigma_5 & \lambda_5^2 \sigma_5^2 \cr
\lambda^1_6  & \lambda^1_6 \sigma_6 & \lambda^1_6 \sigma_6^2
& \lambda_6^2 & \lambda_6^2 \sigma_6 & \lambda_6^2 \sigma_6^2 
}
\pmatrix{
a^2_0 \cr
a^2_1 \cr
a^2_2 \cr
a^1_0 \cr
a^1_1 \cr
a^1_2
}
= 0.
}
A nontrivial solution exists if and only if the determinant
of this matrix is zero:
\eqn\aaa{
0 = 
X = \sum_{i,j,k,l,m,n} \epsilon_{ijklmn} {\cal V}(i,j,k,l,m,n)
\langle i\,l \rangle\langle j\,m \rangle\langle k\,n\rangle.
}
Here ${\cal V}$ is the cyclic product of $\sigma$'s (not the
Vandermonde matrix),
\eqn\xcond{
{\cal V}(i,j,k,l,m,n) = (\sigma_i-\sigma_j)
(\sigma_j - \sigma_k) (\sigma_k-\sigma_l)
(\sigma_l-\sigma_m) (\sigma_m- \sigma_n)
(\sigma_n-\sigma_i).
}
Another way to think about this equation is as follows.
Since one of the $a$'s is fixed by the GL(2) symmetry, we
really only are allowed to solve for five of the $a$'s.
If we choose any five of the equations \five\ to solve
for the five $a$'s and then plug the solution into the sixth
equation, we find the condition that \xcond\ should vanish.

Next we turn our attention to the equations on the second line of
\equations.  These are six ($\dot{\alpha} = 1,2$, $k=0,1,2$)
homogeneous linear equations on the six variables $\xi_i$.
When cast in matrix form, the relevant matrix is precisely
the transpose of \bigone, but with $\lambda \leftrightarrow
\wt$.  A nontrivial solution exists if and only if the
corresponding determinant vanishes:
\eqn\aaa{
0 = \widetilde{X} = 
\sum_{i,j,k,l,m,n} \epsilon_{ijklmn} {\cal V}(i,j,k,l,m,n)
[i\,j] [k\,l] [m\,n].
}

So far we have obtained (subject to $X = 0$) a unique
solution for all of the moduli $a_k^{\dot{\alpha}}$,
and (subject to $\widetilde{X} = 0$) a unique solution for
all of the $\xi_i$.  The final step is to require that
these solutions are compatible, in that they obey
the top line of \equations.
There are a huge number of such compatibility conditions
that one can form, depending on which five of the six equations
\five\ one uses to solve for the moduli and which five of the
six equations from the second line of \equations\ that one
uses to solve for the $\xi_i$.
These equations are polynomials in $\sigma_4$, $\sigma_5$
and $\sigma_6$ whose coefficients depend on $\lambda$, $\wt$
and the fixed values of $\sigma_1$, $\sigma_2$ and $\sigma_3$.

However, these equations
(as well as the $X = 0 = \widetilde{X}$ equations) all
have spurious roots at $\sigma_4 = \sigma_5 = \sigma_6 = \sigma_i$
for $i=1,2,3$.  To eliminate these roots one constructs
a linear combination of these equations (with coefficients involving
powers of $\sigma_4$ and $\sigma_5$), with the coefficients
chosen so that the result factors into a single quartic
polynomial $q(\sigma_6)$ without the spurious roots times
a high-degree polynomial with only spurious roots.

In the previous few paragraphs we have explained in words
the process of constructing a Groebner basis for the ideal
generated by the polynomials \equations.
Once the roots are found, it remains to evaluate the Jacobian.
At the end of the day, the amplitude can be written schematically as
a rational function in $\sigma_6$, summed over $\sigma_6$
satisfying some quartic polynomial:
\eqn\aaa{
A_{6,2} = \sum_{\{\sigma_6 : q(\sigma_6) = 0\}}
{ p(\sigma_6) \over r(\sigma_6)}.
}
Abel's theorem guarantees
that the result of this sum is a rational function of the coefficients
of the polynomials $p$, $q$ and $r$, and it is easy to check
numerically that the result precisely matches the gauge
theory amplitude of Mangano, Parke and Xu \ManganoXK.
More generally, Abel's theorem guarantees that for any $n$ and
$d$, \answer\ turns into a rational function of the covariant
quantities $\langle i\,j\rangle$ and $[i\,j]$ once all of
the roots of \equations\ are summed over.

\subsec{Analysis for special $\lambda$}

Although the $n=6$, $d=2$ amplitude is complicated in
general, instructive analytic expressions can be obtained
by considering special cases.
For example, let us here consider the case
$\lambda_1^2 = \lambda_4^2 = 0$ and $[1\,5] = 0$.
For this degenerate case, numerical investigation reveals
that there are only three roots (one is a double root ---
the statement that $N_{4,2} = 4$ is always true when one counts
multiplicities).  Let us demonstrate
analytically how to find these three roots.

We fix the GL(2) symmetry by setting $a_0^1 = 1$
and $\sigma_i = \{0,1,-1\}$ for $i=1,2,3$.
Also, without loss of generality we
can rescale the $\lambda$'s to set
$\lambda_i^1 = 1$.  From
the $A_i \lambda_i = B_i$ equations for $i=2,3,4,5,6$ we can
solve for the moduli $a^1_1,a^1_2,a^2_1,a^2_2$ and $\sigma_4$ in terms of
$\sigma_5,\sigma_6$.
The first solution is $\sigma_4 = 0$, and the other one is
\eqn\aaa{
\sigma_4=
{{\lambda_3 \lambda_5 \lambda_6 \sigma_{53}\sigma_{63}\sigma_{56} 
+\lambda_2 (2 \lambda_3 \lambda_6 \sigma_5^2
\sigma_{62} \sigma_{63}
}
 {-\lambda_5 \sigma_{52}(\lambda_6 \sigma_{56}\sigma_{63}+
2 \lambda_3 \sigma_{53} \sigma_6^2))}
\over
\lambda_3 \lambda_5 \lambda_6 \sigma_{53}\sigma_{63}\sigma_{56} 
+\lambda_2 ( \lambda_5 \lambda_6 
\sigma_{52}\sigma_{62}\sigma_{56}+
2 \lambda_3 (-\sigma_6 \lambda_5 \sigma_{52} \sigma_{53}+
\lambda_6 \sigma_5 \sigma_{62} \sigma_{63})
}}

The $\sigma_4 = 0$ 
root gives a unique solution for $\sigma_5,\sigma_6$
when we plug the expressions for $a^1_1,a^1_2, a^2_1$ and $a^2_2$
into the equations following from the second line of \equations.
The nonzero $\sigma_4$ root gives a simple solution for $\sigma_6$:
\eqn\aaa{
\sigma_6={[6\,5] (\lambda_3-\lambda_2) \lambda_6 \over
2 [4\,5] \lambda_2 \lambda_3 +[6\,5] (2 \lambda_2 \lambda_3-
\lambda_2 \lambda_6- \lambda_3 \lambda_6)},
}
and a quadratic equation on $\sigma_5$. In other words, the analog of
the fourth order polynomial described in the previous subsection
factorizes into a quadratic one and the square of a linear one.
Solving the equations and plugging them into the Jacobian gives a 
result which agrees numerically with the known gauge theory result.

\newsec{Checks on $n$-Particle Amplitudes}

To summarize, we now know that the formula \answer\ correctly
reproduces all MHV and ${\overline{\rm MHV}}$ amplitudes, as well
as all 6-particle amplitudes.
The nontriviality of these checks makes it implausible
that some complication arises for further amplitudes which
might render \answer\ invalid.
Nevertheless, it would certainly be satisfying to prove
that the formula \answer\ is correct, perhaps  by showing
that it satisfies the recursion relation of \BerendsME.
Since we do not have a complete proof yet, we will content
ourselves with tabulating several consistency checks
that \answer\ is indeed the tree-level $S$-matrix of YM theory for
arbitrary $n$ and $d$.

\subsec{Some properties of gauge theory scattering amplitudes}

Color-ordered
scattering amplitudes in YM theory satisfy a number of important
properties, including:

\noindent
(i) Cyclicity:
\eqn\aaa{
A(2,3,\ldots,n,1) = A(1,2,\ldots,n).
}

\noindent
(ii) Reflection:
\eqn\aaa{
A(n,n-1,\ldots,1) = (-1)^n A(1,2,\ldots,n).
}

\noindent
(iii) Conjugation:
Parity symmetry implies that the amplitude is invariant
under interchanging each helicity $+ \leftrightarrow -$
and simultaneously interchanging $\lambda \leftrightarrow
\wt$,\foot{This transformation makes sense with our choice
of signature (see footnote 1).  In Minkowski signature the left- and
right-hand sides would be related by complex conjugation.}.
The ${\cal{N}} = 4$ supersymmetric version of this statement is
\eqn\conjugation{
A(\lambda_i, \wt_i, \eta_{i A}) = \int d^{4 n} \psi\ \exp\left[
i \sum_{i=1}^n \eta_{i A} \psi_i^A \right]
A(\wt_i,\lambda_i,\psi_i^A).
}

\noindent
(iv) Dual Ward (or Sub-Cyclic) Identity:
\eqn\aaa{
\sum_{C(1,\ldots,n-1)}
A(1,2,3,\ldots,n) =0,
}
where $n$ is held fixed in the last position and
$C(1,\ldots,n-1)$ denotes the set of cyclic
permutations of $\{1,\ldots,n-1\}$.
This identity expresses decoupling of the U(1) degree of freedom
\ManganoBY.

\noindent
(v)
In \BerendsZN\ it was proven that YM amplitudes satisfy the
following generalization of (iv):
\eqn\aaa{
\sum_{{\rm Perm}(i,j)} A(i_1,\ldots,i_m,j_1,\ldots,j_k,n+1) = 0, \qquad
1 \le m \le n - 1, \qquad m + k = n,
}
where the sum is taken over permutations of the set
$(i_1,\ldots,i_m,j_1,\ldots,j_k)$ which preserve the
order of the $(i_1,\ldots,i_m)$ and $(j_1,\ldots,j_k)$ separately.

\noindent
(vi) Soft-Gluon Limit:  In the limit $p_1 \to 0$, the amplitude
behaves as
\eqn\soft{
A(1^+,2,\ldots,n) \longrightarrow
{ \langle n\,2 \rangle \over
\langle n\,1 \rangle\langle 1\,2 \rangle}
A(2,\ldots,n).
}
Of course a conjugated version of this equation should also
hold in the case when particle 1 has negative helicity.   We
do not consider that case directly in this paper, since it
follows as a result of (iii) above.

\noindent
(vii) Collinear Limit:
In the limit $p_1 \to z p$ and $p_2 \to (1-z) p$ for $z \in (0,1)$
and  some
$p$ with $p^2=0$,
the amplitude behaves as
\eqn\collinear{
A(1^+,2^+,3,\ldots,n) \longrightarrow
{1 \over \sqrt{z(1-z)}} {1 \over \langle 1\,2 \rangle}
A(p^+,3,\ldots,n).
}
Again it follows from (iii) that there is an obvious conjugate
to this relation for the case when
particles 1 and 2 both have negative helicity.
The final case, when particles 1 and 2 have opposite helicity, is
\eqn\thirdcollinear{
A(1^+,2^-,3,\ldots,n) \longrightarrow
{z^2 \over \sqrt{z(1-z)}} {1 \over [1\,2]}
A(p^+,3,\ldots,n)
+{(1-z)^2 \over \sqrt{z(1-z)}} {1 \over \langle 1\,2 \rangle}
A(p^-,3,\ldots,n).
}

\noindent
(viii) Multi-particle Poles:
Color-ordered amplitudes can only have poles in channels corresponding
to a sum of cyclically adjacent momenta going on-shell \DixonWI.
That is, if we denote $p_{1,m} = p_1 + p_2 + \cdots + p_m$, then
the amplitude factors in the $p_{1,m}^2 \to 0$ limit according to
\eqn\factorization{
A_n(1,\ldots,n) \longrightarrow
\sum_{\chi = \pm} A_{m+1}(1,\ldots,m,p^\chi)
{i \over p_{1,m}^2} A_{n-m+1}(m+1,\ldots,n,p^{-\chi}).
}

Properties (i), (ii), (iv) and (v) are manifest in \answer\ due to
the way the $\sigma_i$ enter in \measure.
Indeed they follow so trivially from \measure\ that the reader may
well wonder why we have bothered to mention them.
We have done so only because not all of these properties
are immediately obvious from the Feynman diagram expansion of gauge
theory amplitudes.
(These properties are also not all manifest
in the diagrammatic prescription of \CachazoKJ.)

Of the remaining properties, (iii), (vi) and (vii) will be proven
in the following subsections.  The final property (viii)
regarding multi-particle poles
will not be addressed here.  Indeed, note that a proof that
\answer\ satisfies (viii) would essentially be a proof that
\answer\ is correct --- since a tree-level YM amplitude is uniquely
fixed by its poles (and their residues).

\subsec{Parity symmetry}

The parity symmetry \conjugation\ is obvious in gauge
theory but not manifest in the formula \answer\ \foot{The parity
symmetry was also very recently discussed in \BM\ in the framework
of \BerkovitsHG.}.
On the individual
component amplitudes $A_{n,d}$, \conjugation\ says that
\eqn\ccong{
A_{n,d}(\lambda,\wt,\eta) = \int d^{4 n} \psi \ \exp \left[
i \sum_{i=1}^n \eta_{i A} \psi_i^A \right]
\widetilde{A}_{n, n - d - 2} (\wt,\lambda,\psi),
}
thereby relating an integral over the moduli space of
degree $d$ curves to an integral over the moduli space of
degree $n-d-2$ curves.

The proof of \ccong\ is fairly straightforward.
We start by looking for a way to relate the bosonic part of
the amplitudes,
\eqn\rrr{\eqalign{
A_{n,d}(\lambda,\wt) &= \int d {\cal{M}}_{n,d} \prod_{i=1}^n
\delta^2(\lambda_i^\alpha - \xi P_i^\alpha) \prod_{k=0}^d
\delta^2 \left( \sum_{i=1}^n \xi_i \sigma_i^k \wt_i^{\dot{\alpha}}
\right),\cr
\widetilde{A}_{n,n-d-2}(\wt,\lambda) &=
\int d {\cal{\widetilde{M}}}_{n,n-d-2} \prod_{i=1}^n
\delta^2(\wt_i^{\dot{\alpha}} - \widetilde{\xi}
\widetilde{P}_i^{\dot{\alpha}}) \prod_{l=0}^{n-d-2} \delta^2 \left(
\sum_{i=1}^n \widetilde{\xi}_i \widetilde{\sigma}_i^l
\lambda_i^\alpha \right).
}}
Here $d {\cal{\widetilde{M}}}_{n-d-2}$ and
$\widetilde{P}$ are the obvious generalizations of \measure\ and \pdef:
\eqn\aaa{
d {\cal{\widetilde{M}}}_{n-d-2} = 
{d^{2 (n-d-2) + 2} \widetilde{a} \, d^n \widetilde{\sigma}
\, d^n \widetilde{\xi} \over {\rm vol}(GL(2))}
\prod_{i=1}^n {1 \over \widetilde{\xi}_i (\widetilde{\sigma}_i
- \widetilde{\sigma}_{i+1})}, \qquad
\widetilde{P}_i^{\dot{\alpha}} = \sum_{l=0}^{n - d - 2}
\widetilde{a}_l^{\dot{\alpha}} \widetilde{\sigma}_i^l.
}
We will show that after integrating out the moduli $a$,
the first set of delta functions in $A$ exactly transform
into the second set of delta functions in $\widetilde{A}$
(and vice versa) when one makes the change of variables
\eqn\xitdef{
\widetilde{\sigma}_i = \sigma_i, \qquad
\widetilde{\xi}_i = {1 \over \xi_i \prod_{j \ne i} (\sigma_i-\sigma_j)}.
}
The Jacobian for this coordinate transformation is unity,
but we will pick up a simple Jacobian from manipulating the bosonic
delta
functions.  This Jacobian will exactly cancel a similar
fermionic determinant.

Let us begin by studying the quantity
\eqn\pmdef{
p_m = \sum_{i=1}^n { \sigma_i^m \over \prod_{j \ne i}^n
(\sigma_i - \sigma_j)}.
}
We claim that $p_m$ is a polynomial in the $\sigma_i$'s of degree
$m - n + 1$.  To see this, consider $p_m$ as an analytic function
of $z = \sigma_n$ (this can of course be repeated for all of the
$\sigma$'s).
It looks like $p_m(z)$ might have poles
at the other $\sigma_i$, but in fact it is easy to see that the residue
is always zero.  So $p_m(z)$ has no poles, and grows at infinity like
$z^{m - n + 1}$, so it must be a polynomial of degree $m-n+1$.
In particular, $p_m$ vanishes for $m < n - 1$, and $p_{n - 1} = 1$.

Now consider the first type of delta
function in $A$, (we focus on one value of $\alpha$
and restore covariance later)
\eqn\aaa{
I = \int d^{d+1} a \prod_{i=1}^n \delta(\lambda_i - \xi_i P_i),
}
and take linear combinations of the delta functions according
to the $n \times n$ matrix with entries
\eqn\mdef{
M_{mi} = { \sigma_i^m \over \xi_i \prod_{j \ne i} (\sigma_i - \sigma_j)},
\qquad i=1,\ldots,n, \qquad m=0,\ldots,n-1.
}
That is, we write
\eqn\xxx{
I = \int d^{d+1} a \ (\det M) \prod_{m=0}^{n-1} \delta\left(
\sum_{i=1}^n M_{mi} (\lambda_i-\xi_i P_i)\right).
}
The second term in the delta function is now
\eqn\aaa{
\sum_{i=1}^n {\sigma_i^m \over \xi_i \prod_{j \ne i}
(\sigma_i-\sigma_j)} \xi_i \sum_{k=0}^d a_k \sigma_i^k
= \sum_{k=0}^d a_k p_{k+m},
}
using the definitions \pdef, \pmdef\ and \mdef.
Then recalling that $p_{k+m}$ is zero for $m<n-d-1$, we can
split the delta functions into two kinds:
\eqn\aaa{
I = (\det M) \prod_{m=0}^{n-d-2} \delta
\left( \sum_{i=1}^n \widetilde{\xi}_i \sigma_i^m \lambda_i
\right)
\int d^{d+1} a 
\prod_{m=n-d-1}^{n-1} \delta \left(
\sum_{i=1}^n \widetilde{\xi}_i \sigma_i^m\lambda_m
- \sum_{k=0}^d a_k p_{k+m}\right).
}
The $d+1$ moduli now appear linearly in
the last $d+1$ delta functions and can be integrated out
trivially.  The Jacobian for this is just $1$, because
$p_{k+m}$ is a triangular matrix with diagonal entries
$p_{n - 1} = 1$.
Finally, we conclude that
\eqn\yyy{
I = \int d^{d+1} a \prod_{i=1}^n \delta(\lambda_i - \xi_i
P_i) = \left[
V
\prod_{i=1}^n \widetilde{\xi}_i
\right] \prod_{m=0}^{n-d-2} \delta \left( \sum_{i=1}^n \widetilde{\xi}_i
\sigma_i^m \lambda_i\right),
}
where $V$ is the Vandermonde determinant of all of the $\sigma$'s
and the term in brackets comes from evaluating $\det(M)$.

The next step is to simply apply \yyy\ in reverse to get
\eqn\zzz{
\prod_{k=0}^d \delta \left( \sum_{i=1}^n \xi_i \sigma_i^k \wt_i\right)
= \left[ V \prod_{i=1}^n \xi_i \right]^{-1}
 \int d^{(n-d-2) + 1} \widetilde{a}
\prod_{i=1}^n \delta ( \wt_i - \widetilde{\xi}_i \widetilde{P}_i).
}
Finally we can combine \yyy\ and \zzz\ and restore
the $\alpha$ and $\dot{\alpha}$ indices to arrive at
\eqn\aaa{
\eqalign{
&\int d{\cal{M}}_{n,d} \prod_{i=1}^n \delta^2 (\lambda_i^\alpha
- \xi_i P_i^\alpha) \prod_{k=0}^d \delta^2 \left( \sum_{i=1}^n
\xi_i \sigma_i^k \wt_i^{\dot{\alpha}} \right)
\cr
&\qquad\qquad
=
\int d{\cal{\widetilde{M}}}_{n,n-d-2}
\left[ \prod_{i=1}^n
{ \widetilde{\xi}_i \over \xi_i}
\right]^2
\prod_{i=1}^n
\delta^2(\wt_i^{\dot{\alpha}} - \widetilde{\xi} \widetilde{P}_i^\alpha)
\prod_{l=0}^{n-d-2} \delta^2 \left( \sum_{i=1}^n \xi_i \sigma_i^k
\lambda_i^\alpha\right).
}}

We have now related the bosonic integral over degree $d$ curves
to the bosonic integral over degree $n-d-2$ curves, up to a factor
which with the help of \xitdef\ can be written as
\eqn\aaa{
\left[ \prod_{i=1}^n {\widetilde{\xi}_i \over \xi_i}\right]^2
= \left[ V \prod_{i=1}^n\widetilde{\xi}_i \right]^4.
}
In fact, this is precisely the factor which should arise from the
fermionic Fourier transform in the formula \ccong:
\eqn\aaa{
\int d^{4 n} \psi \exp \left[ i \sum_{i=1}^n
\eta_{iA} \psi_{i}^{A} \right] \prod_{l=0}^{n-d-2}
\delta^4 \left( \sum_{i=1}^n \widetilde{\xi}_i \sigma_i^k
\psi_i^A\right)
= \left[ V \prod_{i=1}^n \widetilde{\xi}_i \right]^4
\prod_{k=0}^d \delta^4 \left( \sum_{i=1}^n \xi_i \sigma_i^k
\eta_{i A}\right).
}
This completes the proof that \answer\ satisfies the conjugation
property \ccong.

Incidentally, the above arguments show that given any solution of
the equations \equations\ one can construct a solution of the
conjugate equations
\eqn\ceqn{\eqalign{
\wt_i^{\dot{\alpha}} &= \widetilde{\xi}_i
\sum_{l=0}^{n-d-2} \widetilde{\sigma}_i^l \widetilde{a}_l^{\dot{\alpha}},
\qquad \dot{\alpha} = 1,2, ~~ i=1,\ldots,n,\cr
0 &= \sum_{i=1}^n \widetilde{\xi}_i \widetilde{\sigma}_i^l
\lambda_i^\alpha, \qquad \alpha = 1,2, ~~ l =0,\ldots,n-d-2
}}
by taking $\widetilde{\sigma}$ and $\widetilde{\xi}$ to be
given by \xitdef.
It is not necessary to independently specify the
$\widetilde{a}_l^{\dot{\alpha}}$
since the top equations in \ceqn\ determine them uniquely in terms
of $(\widetilde{\sigma}_i,\widetilde{\xi}_i)$.
Thus, we have shown that
\eqn\duality{
N_{n,n-d-2} = N_{n,d},
}
and moreover, that the contribution to $A_{n,d}$ from any given root
is exactly the conjugate of the contribution of that root to
$A_{n,n-d-2}$.
The relation \duality\ is reminiscent of the relation between
Betti numbers for a manifold of dimension $n-2$ as well as of the
relation between Hodge numbers under mirror symmetry.
It would be interesting to find a relation between $N_{n,d}$ and some
invariants of $\CP^{3|4}$ (perhaps Gromov-Witten invariants)
or of its moduli space of holomorphic curves.

\subsec{The soft gluon limit}

For both the soft gluon and collinear limits, comparing the left-
and right-hand sides of \soft\ and \collinear\ reveals that we
will have to perform two integrals and eliminate two delta functions.
Clearly we want to eliminate the appearance of gluon number 1
on the right-hand side, so we should eliminate the two
delta functions $\delta^2(\lambda_1^a - \xi_1 P_1^\alpha)$
(for $\alpha=1,2$) by performing the integrals over $\xi_1$
and $\sigma_1$.  In general there can be several roots which
contribute to this integral.  However, we are only interested
in roots which in the desired limit give rise to a pole in the
amplitude.
We will argue that only one root contributes to the coefficient
of this pole.

A prototype for both the soft gluon and
collinear limits
involves an integral of the form
\eqn\intone{
I_i = \lim_{\langle 1\, i \rangle \to 0}
\int {d\sigma_1 \over \sigma_1 - \sigma_i}
f(\sigma_1)
\delta \left( 
{\langle i\, 1 \rangle \over \lambda_1^1 \lambda_i^1}
- \left[
{ B_1 \over A_1 } -
{ B_i \over A_i }
\right]
\right).
}
Specifically, we are interested in the
poles of this integral.
We do  not yet need the explicit form of $A$, $B$ or $f$,
and need only to make assumptions which are completely
reasonable for the application at hand:
$B/A$ is a rational function of $\sigma$ with isolated roots,
and the function $f$ has no poles in $\sigma_1$.

The quantity in brackets in \intone\ vanishes when $\sigma_1 = \sigma_i$ and
hence can be written as
\eqn\aaa{
\left[
{ B_1 \over A_1 } -
{ B_i \over A_i }
\right] = (\sigma_1 - \sigma_i) F(\sigma_1 - \sigma_i, \sigma_i)
}
for some $F$.
Changing integration variables from $\sigma_1$ to $w = \sigma_1 - \sigma_i$
gives
\eqn\aaa{
I_i = \lim_{\langle 1\, i\rangle \to 0} \int {dw \over w} \delta\left(
g(w)
\right), \qquad
g(w) = { \langle i\, 1\rangle \over \lambda_1^1 \lambda_i^1}
 - w F(w,\sigma_i).
}
In the limit $\langle 1\, i\rangle \to 0$
the roots of $g(w)$ are easy to analyze.
There is one root (which we will call
$w=w_0$) for which $w$ is small (of the same
order as $\langle 1\, i \rangle$), and there may be
other roots for which $F(w,\sigma_i)$ is small.
We assume there is no degeneracy amongst the possible roots.
Integrating the delta function gives a factor of $1/g'(w)$,
which is a number of order 1 at any of the roots.
Therefore, the only pole in the integral $I_i$ comes from
the factor of $1/w$ evaluated on the root $w=w_0 \to 0$.

The value of $w_0$ is given by the implicit equation
\eqn\ggg{
 w_0 = { \langle i\, 1\rangle \over \lambda_1^1 \lambda_i^1}
{1 \over F(w_0, \sigma_i)},
}
with $F(w_0, \sigma_i)$ being of order unity.
The contribution of this root to the integral is
\eqn\aaa{
I_i = \left. {1 \over w} \left( {\partial g \over \partial w}
\right)^{-1} \right|_{w = w_0}
= {1 \over w_0} \left[ F(w_0, \sigma_i)
+ w_0 \partial_w F(w_0, \sigma_i)\right]^{-1}.
}
Since $F$ is a rational function and $F(w_0, \sigma_i)$
is of order unity, the derivative $\partial_w F(w_0,\sigma_i)$
cannot blow up.  Therefore the second term in brackets can
be ignored as $w_0 \to 0$, so using \ggg\ we arrive at the formula
\eqn\formula{
I_i = \lim_{\langle 1\, i \rangle \to 0}
\int {d\sigma_1 \over \sigma_1 - \sigma_i}
f(\sigma_1)
\delta \left(
{\langle i\, 1 \rangle \over \lambda_1^1 \lambda_i^1}
- \left[
{ B_1 \over A_1 } -
{ B_i \over A_i }
\right]   
\right)
= {\lambda_1^1 \lambda_i^1 \over \langle i\, 1\rangle} f(\sigma_i),
}
which is valid under the assumptions on $f$, $A$ and $B$
given above.

Now let us turn our attention to the soft gluon limit \soft.
First we set the helicity of gluon 1 to $+1$ by setting
$\eta_1 = 0$ in $A_{n,d}$.
This kills the $i=1$ term in the third
delta function in \answer.
In the second delta function, the $i=1$ term also
vanishes trivially in the soft limit since $\wt_1^{\dot{\alpha}} \to 0$.
Particle number 1 therefore only appears in
the integrals
\eqn\aaa{
\int d \sigma_1 \,d\xi_1\ 
{1 \over \xi_1 (\sigma_1 - \sigma_2) (\sigma_n - \sigma_1)}
\delta( \lambda_1^1 - \xi_1 A_1)
\delta(\lambda_1^2 - \xi_1 B_1).
}
The $\xi_1$ integral is trivial and leads to
\eqn\aaa{
{1 \over \sigma_n-\sigma_2}
{1 \over (\lambda_1^1)^2}
\int d\sigma_1 \left[
{1 \over \sigma_1 - \sigma_2} - {1 \over \sigma_1 - \sigma_n}
\right]
\delta\left( {\lambda^2_1 \over \lambda_1^1}
- {B_1 \over A_1}\right).
}
Now we are completely free to subtract from the argument of the
delta function an amount which is equal to zero in the form
of $\lambda^2_i/\lambda^1_i - {B_i/A_i}$ for
any $i \ne 2$.  (This is guaranteed to be zero by the
$\lambda_i^\alpha - \xi_i P^\alpha_i$ delta functions for $i=2$.)
Then we simply apply the formula \formula, once with $i=2$ and
once with $i=n$, to obtain the factor
\eqn\aaa{
{1 \over \sigma_n - \sigma_2} {1 \over (\lambda_1^1)^2}
(I_2 - I_n)
= {1 \over \sigma_n - \sigma_2} {1 \over (\lambda_1^1)^2}
\left[
{ \lambda_1^1 \lambda_2^1
\over \langle 2\, 1\rangle}
- { \lambda_1^1 \lambda_n^1
\over \langle n\, 1\rangle}
\right] = {1 \over \sigma_n - \sigma_2} 
\left[
{ \langle n\,2 \rangle \over 
\langle 2\,1 \rangle \langle n\,1 \rangle}\right].
}
The factor of $1/(\sigma_n - \sigma_2)$ is needed to write
the correct measure factor \measure\ for the $(n-1)$-particle amplitude
$A(2,\ldots,n)$.
Gluon number one has now completely disappeared from the integral,
leaving only the overall factor in brackets, in agreement
with \soft.

\subsec{The collinear limit}

First we consider the
factor
\eqn\jjj{
{1 \over \sigma_n-\sigma_2}
{1 \over (\lambda_1^1)^2}
\int d\sigma_1 \left[
{1 \over \sigma_1 - \sigma_2} - {1 \over \sigma_1 - \sigma_n}
\right]
\delta\left( {\lambda^2_1 \over \lambda_1^1}
- {B_1 \over A_1}\right),
}
which arises exactly as in the previous subsection.
However, whereas we could there use \formula\ for both
$i=2$ and $i=n$ (since $\langle 1\,2\rangle$ and
$\langle 1\,n\rangle$ were both going to zero),
here we can only use \formula\ for $i=2$ since only $\langle 1\,2\rangle
\to 0$ in the collinear limit.
Therefore the second term in brackets in \jjj\  gives
no contribution to the pole, and we only pick up the factor
\eqn\ttt{
{1 \over \sigma_n - \sigma_2} { \lambda_2^1 \over \lambda_1^1}
{1 \over \langle 2 \,1 \rangle} =
{1 \over \sigma_n - \sigma_2}\left[ \sqrt{ 1-z \over z}
{1 \over \langle 2\,1 \rangle}\right].
}

At this stage the integrals over
the variables $\sigma_1$ and $\xi_1$ associated
with gluon number 1 have been performed,
but those associated with
gluon 2 remain and we must rewrite the $\lambda_2$
dependence in terms of $\lambda = \lambda_2/\sqrt{1-z}$.
In the $\xi_2$ integral this is accomplished by
rescaling $\xi_2$ in order to obtain
\eqn\uuu{
\int {d\xi_2 \over \xi_2} \delta(\lambda_2^1 - \xi_2 A_2)
\delta(\lambda_2^2 - \xi_2 B_2) = 
\left[
{1 \over 1-z}
\right]
\int {d\xi'_2 \over \xi'_2} \delta(\lambda^1 - \xi'_2 A_2)
\delta(\lambda^2 - \xi'_2 B_2).
}
The last delta functions to check are the ones of the form
\eqn\aaa{
\eqalign{
\delta^2 \left( \sum_{i = 1}^n \xi_i \sigma_i^k \wt_i^{\dot{\alpha}}
\right) &=
\delta^2 \left( \xi_1 \sigma_1^k \wt_1^{\dot{\alpha}}
+ \xi_2 \sigma_2^k \wt_2^{\dot{\alpha}} +
\sum_{i=3}^n \xi_i \sigma_i^k \wt_i^{\dot{\alpha}}
\right)\cr
&= \delta^2 \left( {\lambda_1^1 \over A_1}
{\sigma_1^k \wt_1^{\dot{\alpha}}
} + \xi_2 \sigma_2^k \wt_2^{\dot{\alpha}}
+ \sum_{i=3}^n \xi_i \sigma_i^k \wt_i^{\dot{\alpha}}
\right)\cr
&= \delta^2 \left( {z \over 1-z} {\lambda_2^1 \over A_2}
{\sigma_2^k \wt_2^{\dot{\alpha}}
} + \xi_2 \sigma_2^k \wt_2^{\dot{\alpha}}
+ \sum_{i=3}^n \xi_i \sigma_i^k \wt_i^{\dot{\alpha}}
\right),
}}
where in the first line we used the fact that we already integrated
out $\xi_1$ setting it to $\xi_1 = \lambda_1^1/A_1$, in the second
line we used the fact that we integrated out $\sigma_1$ setting
$\sigma_1 = \sigma_2$, and in the third line
we used the fact that $\lambda_1 \wt_1 = {z \over (1-z)}
\lambda_2 \wt_2$.
Of course, we know that $\xi_2$ will eventually be set
by a delta function to the value $\lambda_2^1/A_2$, so we may
as well write the final line as
\eqn\aaa{
\delta^2 \left( {1 \over 1 - z} \xi_2 \sigma_2^k \wt_2^{\dot{\alpha}}
+ \sum_{i=3}^n \xi_i \sigma_i^k \wt_i^{\dot{\alpha}}
\right) = 
\delta^2 \left( \xi_2' \sigma_2^k \lambda^{\dot{\alpha}}
+ \sum_{i=3}^n \xi_i \sigma_i^k \wt_i^{\dot{\alpha}}
\right),
}
keeping in mind that $\wt_2 = \sqrt{1-z} \wt$ and
$\xi_2 = \sqrt{1-z} \xi_2'$.

What remains has precisely the structure of
the amplitude $A(p,3,\ldots,n)$, together with the
extra factors in brackets from \ttt\ and \uuu, in complete
agreement with the collinear limit \collinear.
The conjugate of this equation follows from the parity transformation
discussed in section 4.2. The most notable fact following from that
analysis is that the pole arises
from the root satisfying $\sigma_1-\sigma_2\simeq [1\,2]$. 
One might attempt to prove the last collinear limit \thirdcollinear\
by combining the above discussion with this observation. Then, the
different $z$ dependence might arise from the $\lambda$ dependence of the
fermionic integrals.

\newsec{Conclusions and Speculations}

In this paper we have
presented strong evidence that the formula \answer\ encodes
the complete tree-level $S$-matrix of Yang-Mills theory in four
dimensions.
Explicit calculation has now shown that \answer\ agrees with YM
theory for all MHV and ${\overline{\rm MHV}}$ amplitudes, as well
as all 6-particle non-MHV amplitudes.  Moreover the analysis of
section 4 shows that for any $n$, \answer\ satisfies a number
of important properties required of gauge theory amplitudes,
including parity symmetry.
Many interesting directions remain open.

Of primary importance is to understand the connection between
the formula \answer,
which was obtained in \RoibanVT\ following the suggestion in
\WittenNN\ that one should consider a single instanton of
degree $d$ in the topological B-model on $\CP^{3|4}$, and
the diagrammatic procedure of \CachazoKJ, in which arbitrary
amplitudes are built out of $d$ disconnected amplitudes, each
of degree 1.  We suspect that formulating a proof that
\answer\ factorizes correctly onto multi-particle
poles would essentially amount to proving the equivalence
of \answer\ and the rules of \CachazoKJ, simply because
the factorization properties are completely manifest in the latter.

The numerical coefficient in front of \answer\ was fixed by
comparing with gauge theory.  We have not computed this
coefficient independently in the B-model.  It is conceivable
that the degree $d$ contribution and the separated degree $1$
contributions (as well as other contributions) have to be
added together to fully reproduce the 
normalization of the
gauge theory scattering amplitudes.
It is also possible that the B-model
has some huge symmetry group 
which relates
the connected instanton contribution
\answer\ 
to the fully disconnected instantons of \CachazoKJ.

Of course, even forgetting for the moment about the
B-model, it would also be very interesting
to prove rigorously
that the formula \answer\ is the tree-level $S$-matrix
of Yang-Mills theory.
To this end it would be useful to understand
better the mathematical structure of the equations \equations,
and in particular to learn how many roots they have for
general $n$ and $d$ (i.e., what is the degree
of the corresponding Groebner basis).  These numbers might be related
to some interesting invariants
of $\CP^{3|4}$ or of its moduli space of holomorphic curves, and perhaps
the equality of $N_{n,d}$ and $N_{n,n-d-2}$ could be understood in this
language.

Finally, all of our considerations have applied to the tree-level
S-matrix in gauge theory.
An obvious next
step of great interest would be to see what light the topological B-model
can shed on one-loop calculations \BernJE.

\bigbreak\bigskip\bigskip\centerline{\bf Acknowledgments}\nobreak

We have benefited from helpful discussions
with D.~Berenstein, Z.~Bern, D.~Gross, C.~Herzog, D.~Kosower, L.~Motl and
E.~Witten.
This
research was supported in part by the National Science Foundation under
Grant No.~PHY99-07949 (MS, AV) and PHY00-98395 (RR), as well as
by the DOE under Grant
No.~91ER40618 (RR).  Any opinions, findings, and conclusions or
recommendations expressed in this material are those of the authors
and do not necessarily reflect the views of the National Science
Foundation.

\listrefs

\end